\documentclass[twocolumn]{aastex62}
\usepackage{bm}
\usepackage{graphicx}
\usepackage{wasysym}
\begin{document}

\title{\textbf{\Large{Modeling Saturn's D68 clumps as a co-orbital satellite system \\}}}

\author[0000-0001-7522-7806]{Joseph A. A'Hearn}
\affiliation{Department of Physics, University of Idaho, Moscow, Idaho, USA}

\author[0000-0002-8592-0812]{Matthew M. Hedman}
\affiliation{Department of Physics, University of Idaho, Moscow, Idaho, USA}

\author[0000-0002-5010-0574]{Douglas P. Hamilton}
\affiliation{Department of Astronomy, University of Maryland, College Park, Maryland, USA}

\correspondingauthor{Joseph A'Hearn}
\email{jahearn@uidaho.edu}

\begin{abstract}
The D68 ringlet is the innermost feature in Saturn's rings. 
Four clumps that appeared in D68 around 2014 remained evenly spaced about 30$^{\circ}$ apart 
and moved very slowly relative to each other from 2014 up until the last measurements were taken in 2017. 
D68's narrowness and the distribution of clumps could 
either indicate that we have a collection of source bodies in a co-orbital configuration
or imply that an outside force confines the observed dust and any source bodies. 
In this paper we explore the possibility that these four clumps arose from four source bodies in a co-orbital configuration. 
We find that there are no solutions with four masses that produce the observed spacings.
We therefore consider whether an unseen fifth co-orbital object could account for the discrepancies in the angular separations 
and approach a stable stationary configuration. 
We find a range of solutions for five co-orbital objects where their mass ratios depend on the assumed location of the fifth mass.
Numerical simulations of five co-orbitals are highly sensitive to initial conditions, 
especially for the range of masses we would expect the D68 clumps to have. 
The fragility of our D68 co-orbital system model implies that there is probably some outside force confining the material in this ringlet. 
\\
\end{abstract} 

% There is a new list of keywords being used these days.
%\keywords{planets and satellites: rings} %there is a list of keywords on the aj/apj website

%\keywords{Unified Astronomy Thesaurus Concepts: dynamical evolution (421), orbits (1184), planetary rings (1254)}

\section{Introduction: Four long-lived bright clumps in the D ring \label{intro}}
A narrow ringlet referred to as D68 lies near the inner edge of Saturn's D ring, about 67,630 km from Saturn's center.
From its discovery in Voyager images \citep{Showalter96} through much of the Cassini mission, 
investigation of D68 focused on its radial profile and phase angle properties \citep{Hedman07}. 
Later studies brought attention to its longitudinal brightness variations \citep{Hedman14}. 
In 2014-15, four bright clumps %(designated T, M, L, and LL; see Figure \ref{clump}) 
formed and remained relatively evenly spaced 
with small longitudinal variations about mean separations of 26$^{\circ}$, 32$^{\circ}$, and 29$^{\circ}$ \citep{Hedman19}. 
\citet{Hedman19} investigated these clumps in depth and designated them T (trailing), M (middle), L (leading), and LL (leading leading). 
The most likely explanation for the sudden increase in brightness in the four clump regions of the ringlet is that fine material was released by collisions into or among larger objects located near or within D68. 
These hypothetical larger objects are called source bodies, whose minimum sizes can be constrained by estimating the amount of material associated with each clump from phase-corrected normal equivalent area values, and whose maximum sizes can be constrained by the fact that they have not been observed directly. 
The range of masses that would correspond to these size constraints is $10^5-10^{10}$ kg. 
The narrowness of the D68 ringlet and the distribution of clumps could 
either indicate that there is a collection of source bodies in a co-orbital configuration
or imply that there is some outside force confining this material. 
In this paper we test the first idea by modeling the D68 clumps as a co-orbital satellite system.  

The study of the dynamics of co-orbital systems is motivated by the many cases of co-orbital systems we find in our solar system. 
We are especially interested here in systems in which the co-orbitals have comparable masses. 
The best known of such systems are the horseshoe orbits of Janus and Epimetheus \citep{Dermott81}.
Co-orbital asteroids have been suggested as the source of Venus's zodiacal dust ring \citep{Pokorny19}.
Finally, the ring arcs in the Neptunian system have been proposed to be confined by either a corotation resonance with a moon on a separate orbit \citep{Goldreich86, Porco91, Salo98, Namouni02} or a co-orbital resonance with an undetected moon or even multiple moons sharing the same orbit \citep{Lissauer85, Sicardy92, Renner14}.  

%\begin{figure}[h]
%\centerline{\includegraphics[width=\columnwidth]{clump_example_high_res}}
%\caption{This image, taken on 2 February 2016, shows one of the bright clumps in the D68 ringlet.
%D68 is the narrow ring near the center of the image. The bright feature is located near the ansa. \citep{Hedman19}
%}
%\label{clump}
%\end{figure}

In Section \ref{analysis}, we analyze potential stable configurations. 
In Section \ref{numerical}, we describe how we use numerical simulations to investigate these scenarios.
In Section \ref{disc}, we discuss some remarks for co-orbital systems as well as the possibilities for D68. 

\section{Analysis of Potential Stable Configurations \label{analysis}}
Here we first review the theory of stable co-orbital objects, 
and we then apply the theory to the D68 clumps.

\subsection{Theory \label{theory}}

\citet{Salo88} originally examined stationary configurations of equal-mass co-orbital satellites for small $N$ ($N \leq 9$)
using a simple first-order theory, neglecting terms of the order $\left(m/M\right)^{3/2}$, 
where $m$ and $M$ are the masses of the satellite and the primary. 
A numerical search revealed three distinct types of stationary solutions, 
of which we are here concerned with only one, which \citet{Salo88} label Type Ia: 
an equilibrium where all the $N$ satellites are most concentrated on the same side of the common orbit. 
The case where $N = 2$ is the well known Trojan configuration, with an angular separation of 60$^{\circ}$. 
Type Ia configurations are stable for $2 \leq N \leq 8$ but are not found for $N \geq 9$ \citep{Salo88}.
This study, motivated by the D68 clumps, focuses on configurations with $N = 4$ and $N = 5$. 
\citet{Renner04} generalized the work of \citet{Salo88} for similar but not necessarily equal masses, 
which is what we expect for the D68 clumps. 

When we define $\phi_i$ as the longitude of satellite $i$ and $\xi_i = \Delta r_i/r_0$ as its relative radial excursion with respect to its average radius $r_0$,
the relevant equations of motion become \citep{Renner04}
\begin{equation}
\dot{\phi}_i = -\frac{3}{2}\xi_i
\end{equation}
and
\begin{equation}
\dot{\xi}_i = -2 \displaystyle\sum_{j \neq i} m_j f'\left(\phi_i - \phi_j\right)
\end{equation}
where $m_j$ is the mass of satellite $j$ and
\begin{equation}
f\left(\phi\right) = \cos \phi - \frac{1}{2|\sin\frac{\phi}{2}|}.
\end{equation}
The derivative of $f\left(\phi\right)$ is 
\begin{equation}
f'\left(\phi\right) = \sin \phi \left[-1 + \frac{1}{8|\sin\frac{\phi}{2}|^3} \right].
\end{equation}
For a co-orbital stationary configuration \citep{Renner04}, 
\begin{equation}
\xi = 0 
\end{equation}
and 
\begin{equation}
\displaystyle\sum_{j \neq i} m_j f'\left(\phi_i - \phi_j\right) = 0.
\label{mf}
\end{equation}
Equation \ref{mf} can be written in matrix form:
\begin{equation}
\pmatrix{ 0 & f'_{12} & \dots & \dots & f'_{1N} \cr
                -f'_{12}       & 0 & f'_{23} & \dots & f'_{2N} \cr
                \vdots & & 0 & & \cr
                \vdots & & & \ddots & \cr
                -f'_{1N} &  &  & & 0 }
\pmatrix{ m_1 \cr
               m_2 \cr
               \vdots \cr
               \vdots \cr
               m_N }
= 0_{\rm I\!R^N}
\end{equation}
Because the $N \times N$ matrix is antisymmetric and depends only on the longitudinal separations $\phi_i$ between the bodies, 
for $N \ge 3$ one can always find a set of relative masses that satisfies these equations
for any given set of angular separations. 
This solution, however, might not be physical because one or more of the masses could be negative or zero.

\subsection{Results \label{results1}}
We first considered the observed configuration with four masses separated by angles of 29$^{\circ}$, 32$^{\circ}$, and 26$^{\circ}$ because these are the observed separations \citep{Hedman19}.
These separations are closer than the expected separations for an equal-mass situation: 41.498$^{\circ}$, 37.356$^{\circ}$, and 41.498$^{\circ}$ \citep{Salo88}.
We therefore solved the above equations for arbitrary masses, using Gaussian elimination, which involved re-ordering the rows, and found that the solution contains a mass $\epsilon$ that is calculated as either zero or a small negative number on the order of $10^{-16}-10^{-19}$, 
depending on the order in which the rows are solved (most likely a numerical issue involving the limit of double precision numbers):
\begin{equation}
\pmatrix{ m_1 \cr
               m_2 \cr
               m_3 \cr
               m_4 }
= \pmatrix{ m_T \cr
               m_M \cr
               m_L \cr
               m_{LL} }
= \pmatrix{ \epsilon \cr
               0.55825 \cr
               0.00635 \cr
               0.4354 }
\end{equation}
when we normalize the relative masses such that their sum is 1.
%Thus, there does not exist a physical solution for mass ratios of the four clumps in a stationary configuration. 
Thus, there does not exist a physical solution for the stationary configuration with four objects that would produce four comparably bright clumps.

There are two possible ways that the clumps could still reflect a collection of co-orbital source bodies: 
the four source bodies could have been librating around the equilibrium location 
or there could be another massive body in the system that did not produce a visible clump.
It is certainly possible for there to be only four non-stationary clumps and for this to be a transient phenomenon. 
In fact, \citet{Hedman19} 
identified slow changes in the clumps' azimuthal separations over time that could be evidence for libration.
It is unlikely, however, for the clumps to be on the edge of a libration cycle, 
due to how azimuthally compact the whole configuration is. 
The most compact state of a configuration of three is in the symmetric mode when the outer bodies are at their closest approach to the middle body. 
A similar symmetric mode in a system of four bodies would require the outer two bodies to converge at a faster rate than the middle two bodies.
The observed drift rates, however, show the opposite trend, with the middle two clumps drifting at a faster rate than the outer clumps \citep{Hedman19}.

\begin{figure}[h]
\centerline{\includegraphics[width=0.7\columnwidth]{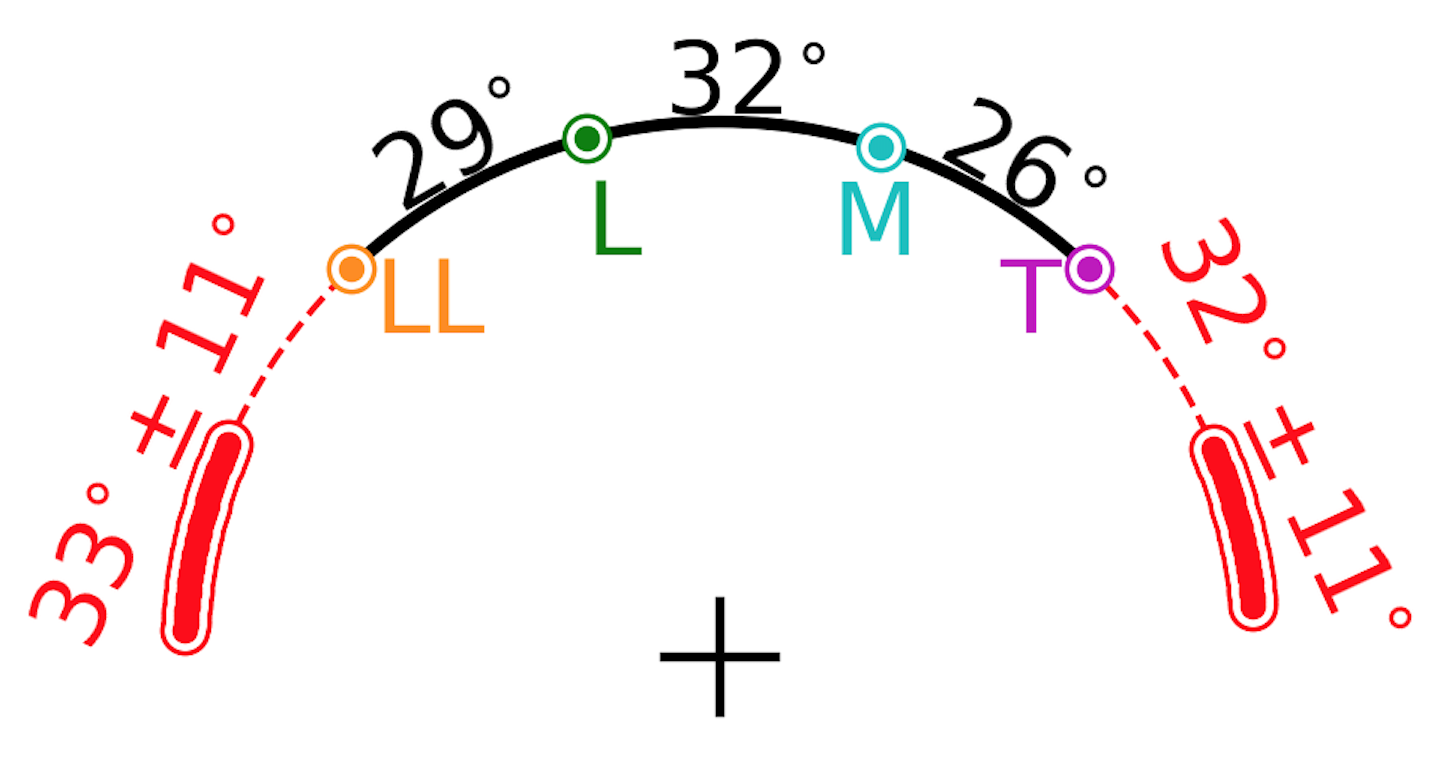}}
\caption{The configuration of the four D68 clumps along with the two regions where a fifth object could be. 
One region is leading, the other is trailing. 
The direction of orbital motion is counterclockwise.
}
\label{D68p5as}
\end{figure}
\begin{figure}[h]
\centerline{\includegraphics[width=\columnwidth]{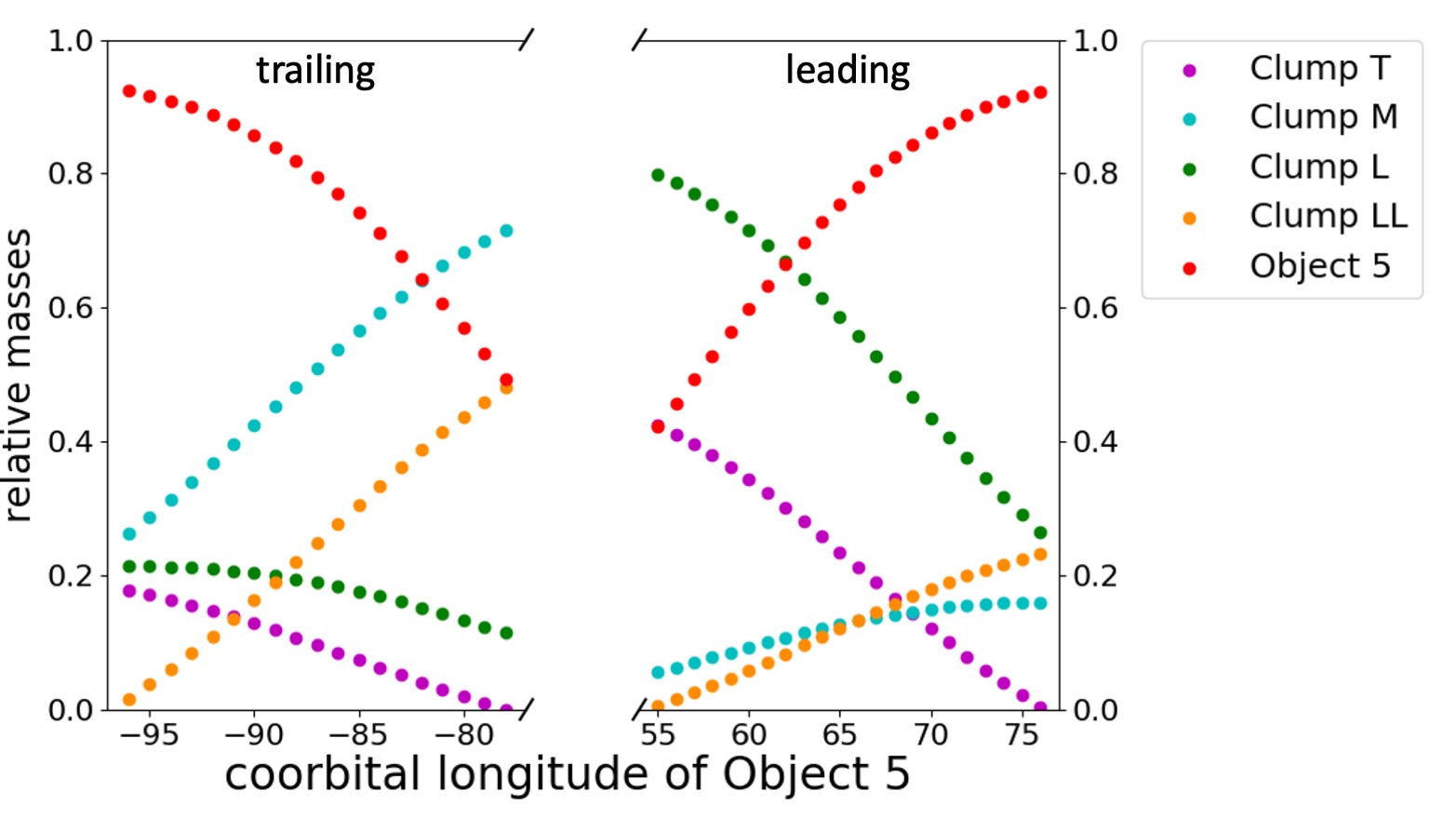}}
\caption{This plot shows the relative masses of the five co-orbitals for each possible configuration. 
Compact configurations are characterized by the more massive bodies in positions 1, 3, and 5. 
Extended configurations, by contrast, have the most massive object on one end, with the other masses tending to decrease with increasing distance.
}
\label{rel_mass}
\end{figure}

If, however, the dust around four source bodies was stirred up by an object that passed nearby, 
it is certainly possible this object could have missed other source bodies in the D68 ringlet.
We therefore consider whether there could be an unseen fifth object, 
whose mass could account for the angular separations we observe between the four known clumps. 
We explore the approximately 270-degree span of longitudes ahead of Clump LL and behind Clump T.
Using the same equations of motion \citep{Renner04}, we find physically realistic solutions in two regions, 
one centered 33$^{\circ}$ ahead of Clump LL and one centered 32$^{\circ}$ behind Clump T. 
These regions each span about 22$^{\circ}$ in longitude and are mapped out in Figure \ref{D68p5as}.
The relative masses of the clumps that correspond to these solutions are plotted in Figure \ref{rel_mass}.
The horizontal axis shows the longitude of Object 5 in the same longitude reference system used by \citet{Hedman19}.
The left-hand side of the split horizontal axis corresponds to a configuration in which Object 5 is trailing the other D68 clumps; 
the right-hand side corresponds to a configuration in which Object 5 is leading the other D68 clumps. 
In more compact configurations (when Object 5 is near longitudes -80 and 55), the middle and outer masses are greater than the second and fourth masses.  
In less compact configurations (when Object 5 is near longitudes -95 and 75), Object 5's mass would be more than double that of any other mass, 
and the clump farthest away from Object 5 becomes the least massive while the other three would require comparable masses. 

\section{Numerical Investigations of the Configurations' Stability \label{numerical}}
To further examine the dynamics of a co-orbital system at the semi-major axis of the D68 ringlet and to investigate stability limits, 
we numerically simulated the motion of point masses at the longitudes of the clump peaks, 
adding in a fifth point mass at one of the locations permitted by the methods found in \citet{Renner04}.  
For orbital simulations, we used the hybrid symplectic/Bulirsch-Stoer algorithm in the Mercury6 package \citep{Chambers99}. 
Our orbital simulations considered Saturn as the central mass and included terms up to $J_6$ in its gravitational field. 
The constants used for these simulations were taken from \citet{Jacobson06} and \citet{Archinal18}, 
converted to the units used in Mercury6, and are found in Table \ref{tbl-1}. 
We used a time step of 0.02 days, which for D68 corresponds to about one tenth of an orbit. 

\begin{table}
\begin{center}
\caption{Parameters of Saturn used for numerical simulations, from \citet{Archinal18} and \citet{Jacobson06} \label{tbl-1}}
\begin{tabular}{crr}
\tableline
\tableline
Parameter & Value \\
\tableline
R$_{\saturn}$ &60,268 km \\
GM$_{\saturn}$ &37931207.7 km$^3$ s$^{-2}$\\
$J_2$ &1.629071$\times 10^{-2}$\\
$J_4$ &-9.3583$\times 10^{-4}$ \\
$J_6$ &8.614$\times 10^{-5}$ \\
\tableline
\end{tabular}
\end{center}
\end{table}

For the sake of simplicity, we focused on one specific stable solution with the corresponding angular separation of Object 5 in order to do numerical simulations,
though other configurations were also investigated, both on the leading and trailing sides, to ensure that our conclusions are general. 
We focus on a configuration with Object 5 ahead of Clump LL by 33$^{\circ}$, as specified in Table \ref{tbl-2}. 
\begin{table}
\begin{center}
\caption{Initial parameters of co-orbitals used for numerical simulations \label{tbl-2}}
\begin{tabular}{crrr}
\tableline
$a$ &67,627 km \\
\tableline
\tableline
Body & Mean longitude & Relative mass &\\
\tableline
T &$122^{\circ}$ &0.129 &\\
M &$148^{\circ}$ &0.069 &\\
L &$180^{\circ}$ &0.322 &\\
LL &$209^{\circ}$ &0.066 &\\
5 &$242^{\circ}$ &0.414 &\\
\tableline
\end{tabular}
\medskip
\\ Note: Relative mass is normalized such that the sum of all five masses equals 1. 
\end{center}
\end{table}

We explored perturbations to this configuration in semi-major axis and longitude, 
modifying the initial semi-major axis or longitude for some of the bodies.
We also varied their absolute mass, while keeping their relative masses constant, as calculated above \citep{Renner04}. 
Although the highest mass range we expect for the clump source bodies is $10^{9}-10^{10}$ kg because they have not been observed directly \citep{Hedman19}, 
we also consider much more massive configurations because these evolve more quickly 
and in this way clarify how these systems respond to perturbations.
Thus we consider three different situations: one with extreme masses of $10^{20}-10^{21}$ kg (i.e., similar to Enceladus, Tethys, and Dione), 
one with with masses of $10^{13}-10^{14}$ kg (i.e., similar to Polydeuces, Pallene, and Daphnis), 
and one with masses of $10^{8}-10^{9}$ kg,  close to that expected for the D68 source bodies.

In each simulation, we plot the longitudinal evolution of the bodies with respect to a reference longitude, which is calculated for each timestep as
\begin{equation}
\lambda_0 = \arctan{\frac{\sum_i^N{\sin{\lambda_i}}}{\sum_i^N{\cos{\lambda_i}}}}
\end{equation}
where $\lambda_i$ is the mean longitude of body $i$. 
This equation works well when the longitudinal oscillations are small. 
This type of plot gives a quick sense of stability and of orbital evolution.

\begin{figure}[h]
\centerline{\includegraphics[width=\columnwidth]{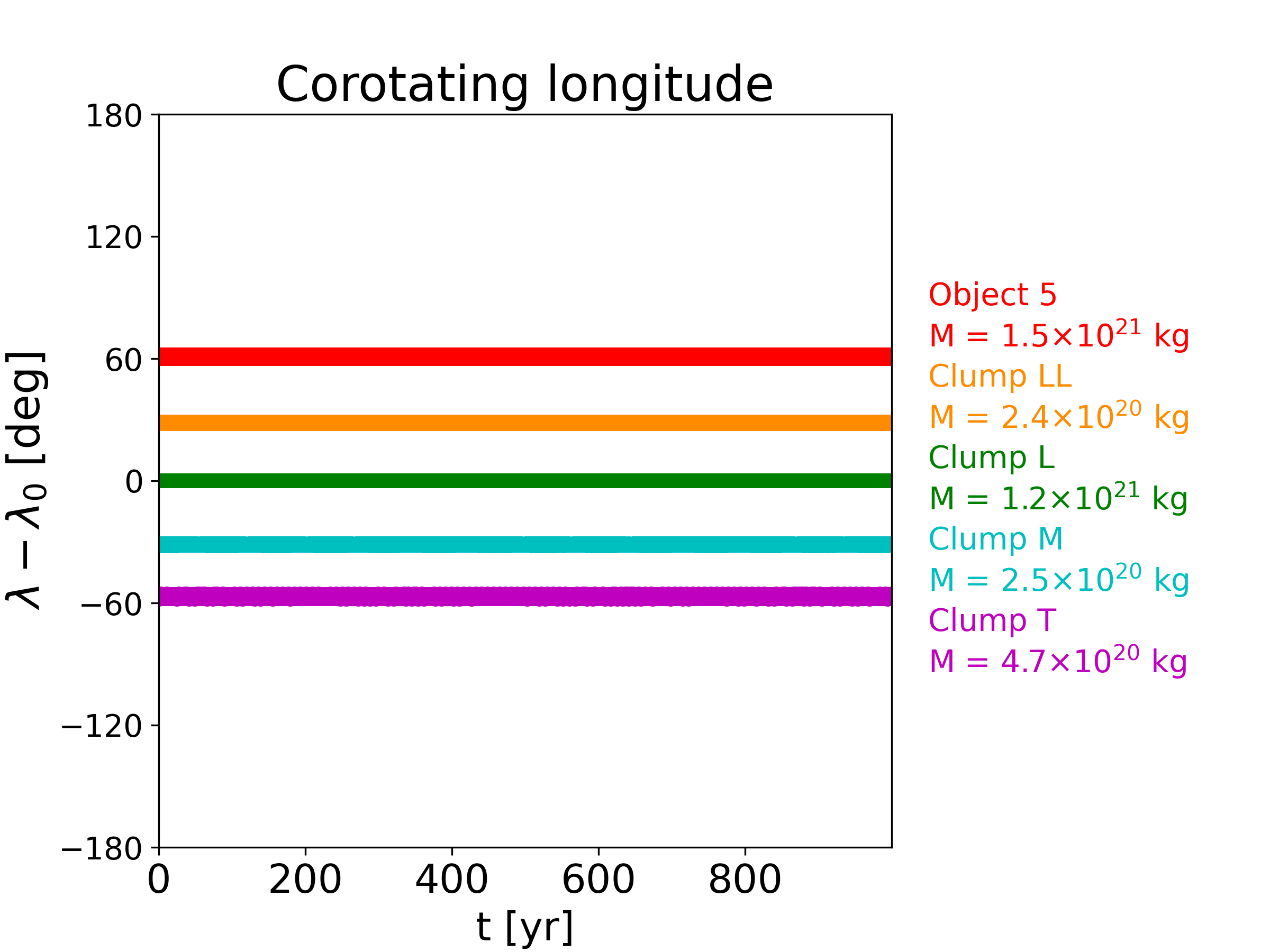}}
\caption{With extremely high masses and no perturbations
(initial angular separations of 33$^{\circ}$, 29$^{\circ}$, 32$^{\circ}$, and 26$^{\circ}$), 
the system is stable, consistent with the analytic theory.
}
\label{large0}
\end{figure}

We verify that the stationary points found using the method of \citet{Renner04} are indeed stable 
by placing objects there and finding they do not evolve in 1000-year simulations with high masses ($10^{20}-10^{21}$ kg; see Figure \ref{large0}).
Here we do not explore perturbations in initial longitude or semi-major axis for the high-mass case
because the larger masses complicate scalings to the real system.  

\begin{figure}[h]
\centerline{\includegraphics[width=\columnwidth]{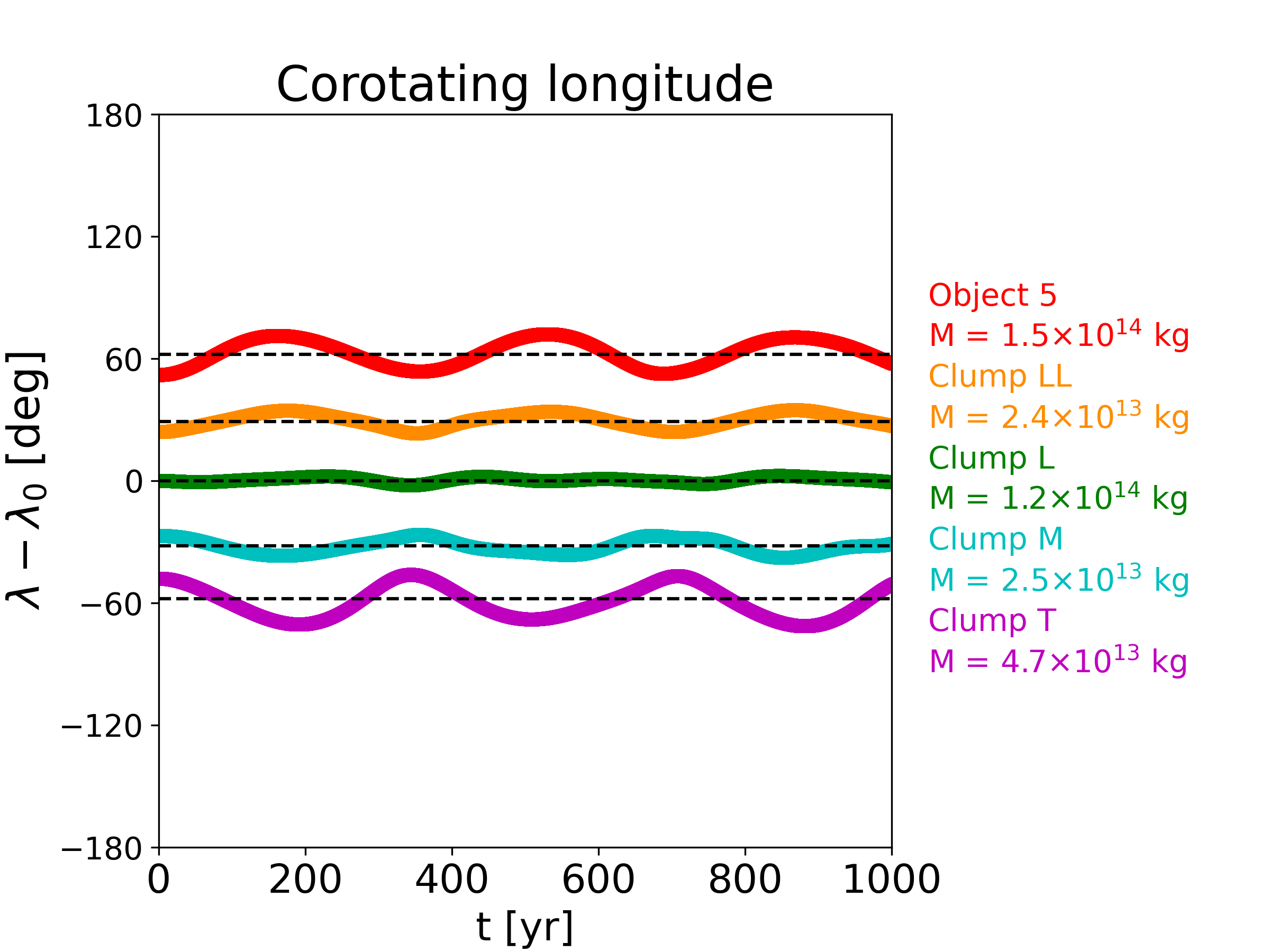}}
\caption{Including an initial perturbation to longitudes to start the system in a more compact configuration
(with initial angular separations of 28$^{\circ}$, 24$^{\circ}$, 27$^{\circ}$, and 21$^{\circ}$), 
the bodies oscillate around the stable solution, which is indicated by the dashed lines.
}
\label{compact}
\end{figure}

We consider two types of perturbation, longitudinal and radial, in the medium-mass case,
$10^{13}-10^{14}$ kg.
The objects are massive enough that it is easier to demonstrate both stable libration and more chaotic mutual encounters.
First, we consider a longitudinal shift in which the system begins in a more compact configuration, 
and we find that the masses oscillate stably around the solution (see Figure \ref{compact}).

\begin{figure}[h]
\centerline{\includegraphics[width=\columnwidth]{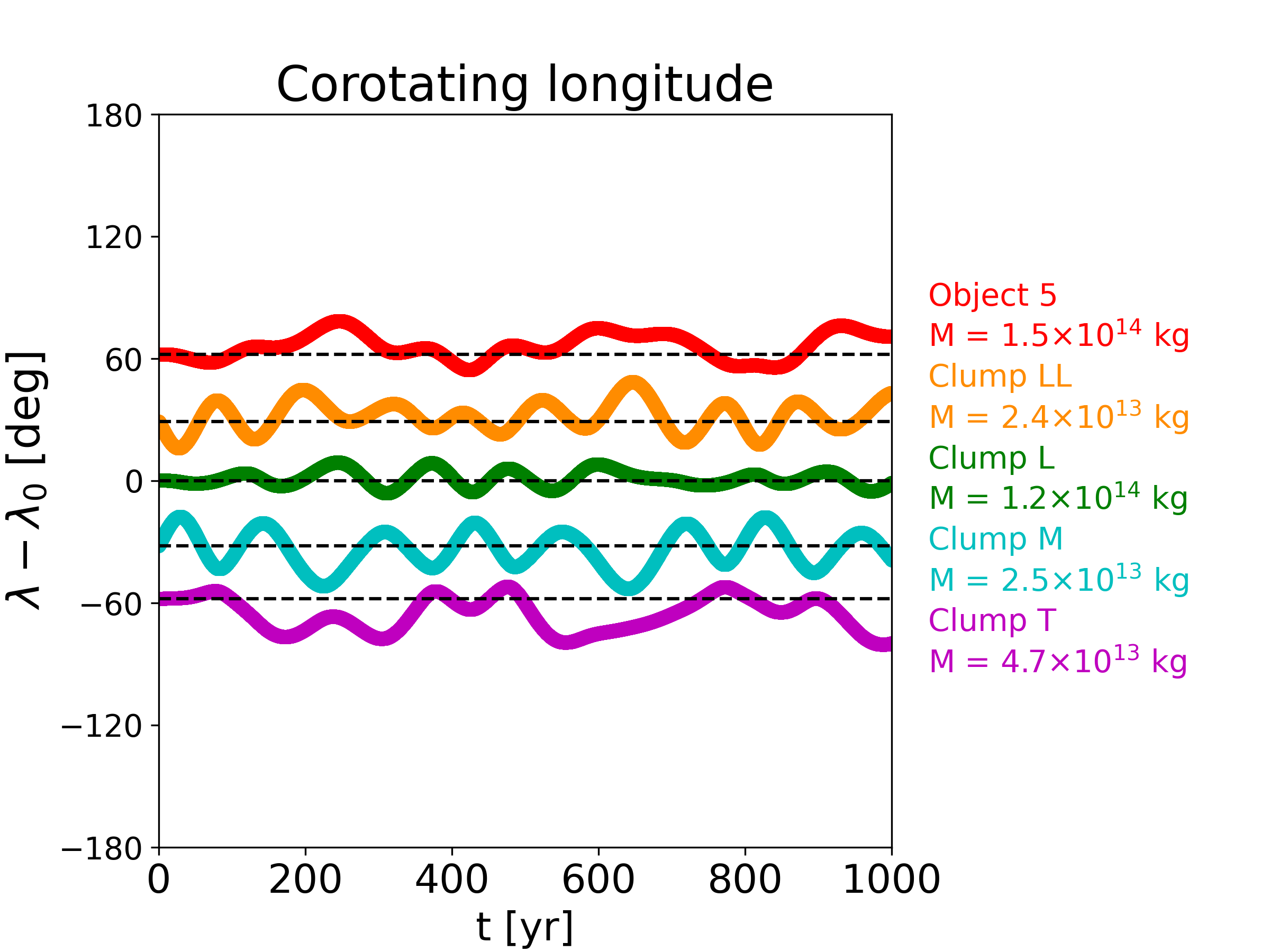}}
\caption{With sufficiently small initial perturbations to semi-major axes (50 m for Daphnis-scale co-orbitals), 
the bodies oscillate around a stable solution, which is indicated by the dashed lines.
}
\label{med50}
\end{figure}

\begin{figure}[h]
\centerline{\includegraphics[width=\columnwidth]{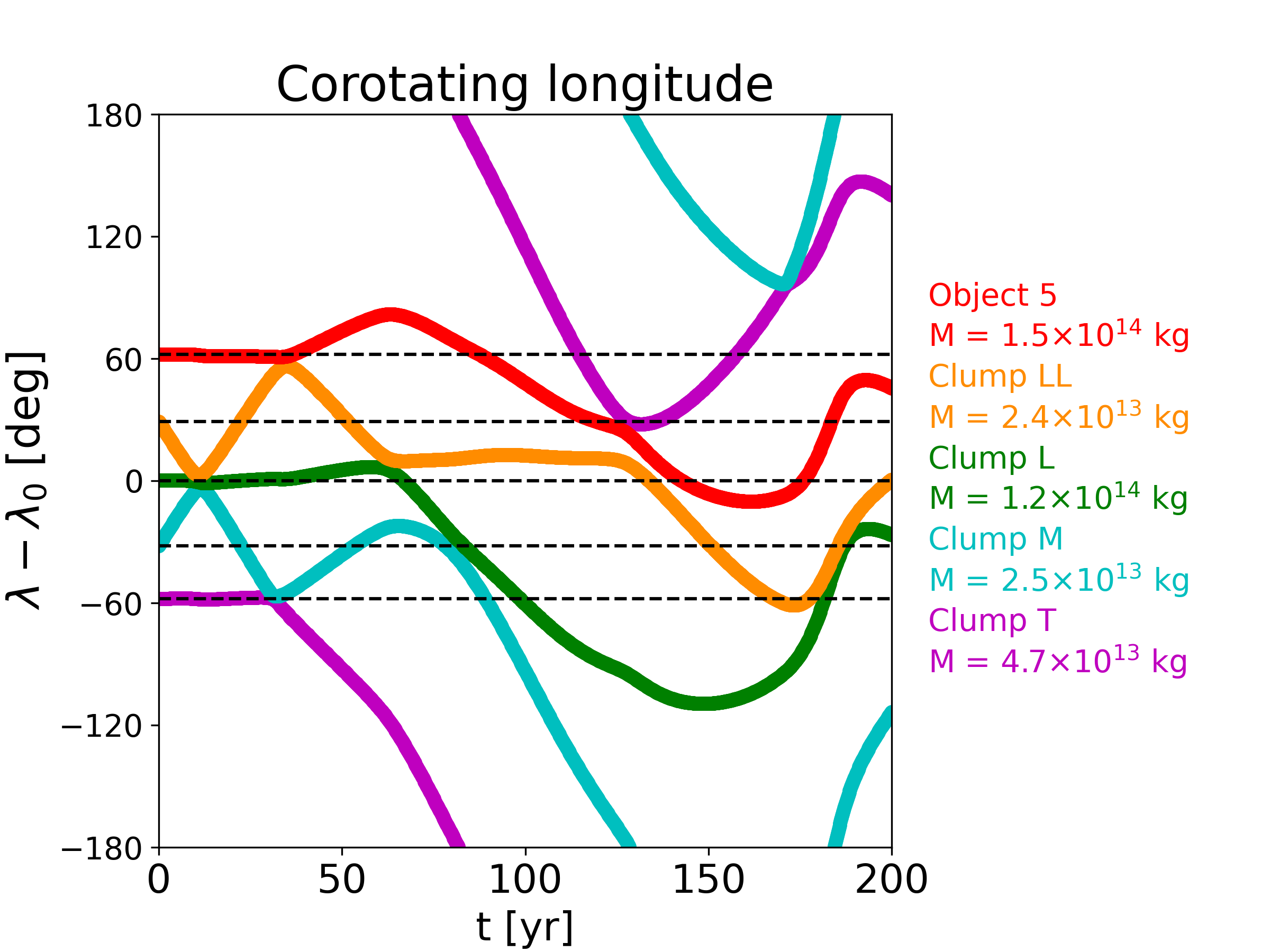}}
\caption{With large enough initial perturbations to semi-major axes (200 m for Daphnis-scale co-orbitals), 
the system becomes unstable when some of the bodies encounter each other.
}
\label{med500}
\end{figure}

Second, we consider radial perturbations in which we modify the initial semi-major axis. 
We define a critical semi-major axis separation $\Delta a_{\rm{crit}}$ which separates 
small oscillatory motion like that shown in Figure \ref{med50} from the sort of motion shown in Figure \ref{med500}.
We explored through simulations the allowable perturbations to semi-major axis using the mode in which 
Clump LL is given a positive $\Delta a$ and Clump M is given a negative $\Delta a$, just as in Figure \ref{med50}. 
We found that, for these relative masses in this specific perturbation mode, the critical semi-major axis separation's relation to absolute mass is best represented as
\begin{equation}
\frac{\Delta a_{\rm{crit}}}{a} \simeq 1.06 \left(\frac{m_{\rm{clumps}}}{M_{\rm{planet}}}\right)^{0.49}
\end{equation}
%We expect that the coefficient of this equation and perhaps even the exponent could vary for different perturbation modes and different numbers of co-orbitals, and though that is beyond the scope of this project, we hope to investigate this in the future. 
For the medium-mass case, $\Delta a_{\rm{crit}} = 75.8$ m, which occurs in between the cases 
shown in Figures \ref{med50} and \ref{med500}, namely, 50 m and 200 m. 
Perturbations of $\Delta a = 50$ m
are small enough that when two of the bodies approach each other, 
they exchange energy and angular momentum in such a way as to begin receding from each other, similar to the periodic orbital swap of Janus and Epimetheus. 
Perturbations of $\Delta a = 200$ m
are too much for a stable configuration, which results in bodies looping around to approach the other side of the co-orbital system 
and eventual spreading into multiple orbits via gravitational interactions with the other bodies. 

\begin{figure}[h]
\centerline{\includegraphics[width=\columnwidth]{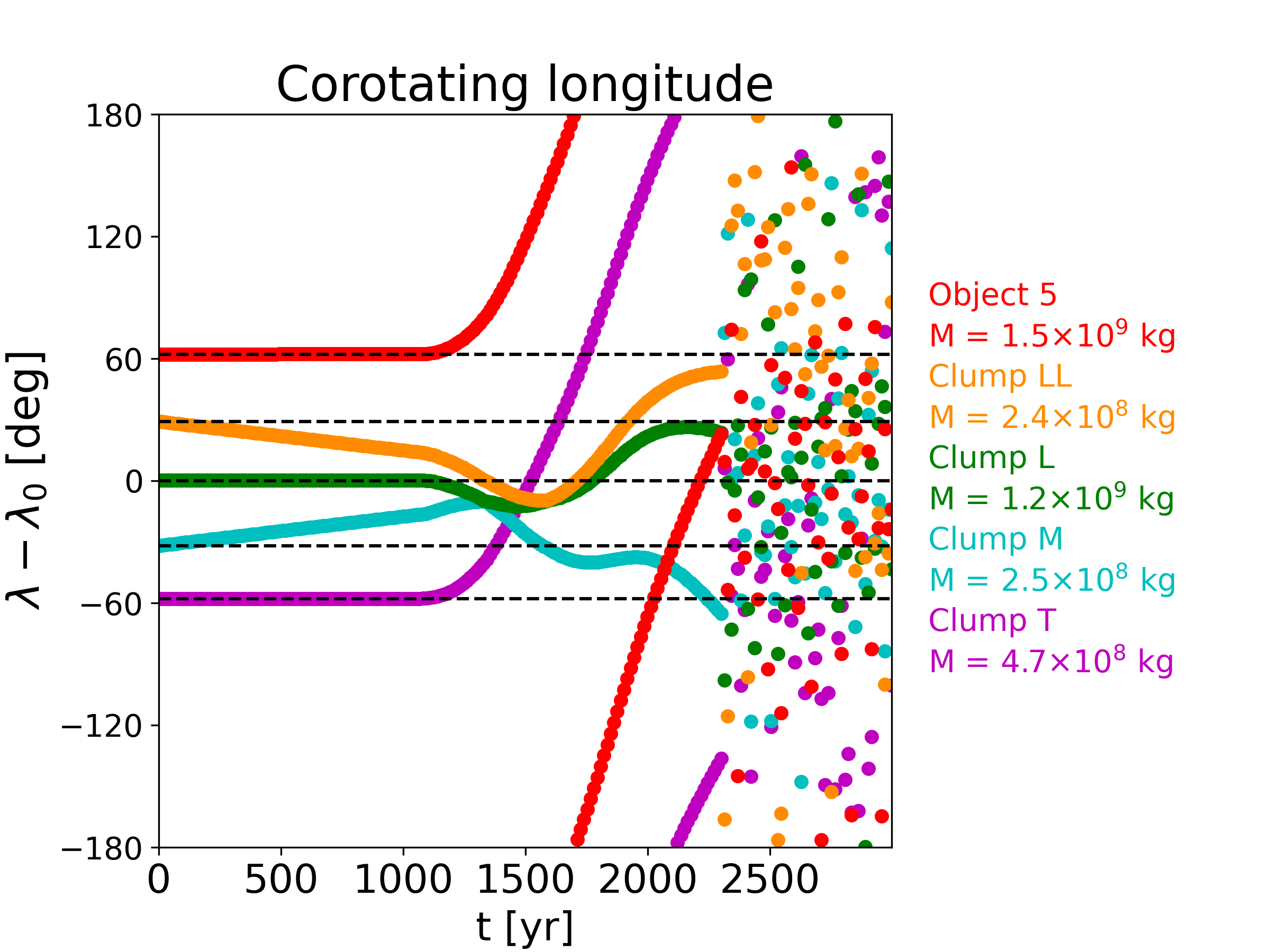}}
\caption{With realistic masses, 
even semi-major axis perturbations of one meter result in system instability. 
Although low-mass co-orbital systems are fragile, stability could be achieved with the help of an external force.
% for example, dynamical friction with a sea of smaller bodies.
}
\label{low1}
\end{figure}

To apply our numerical simulations to the D68 clumps, however, we must also consider the dynamics in a low-mass case,
$10^{8}-10^{9}$ kg.
For the low-mass case, $\Delta a_{\rm{crit}} = 27$ cm, which is confirmed by Figure \ref{low1}.
With only 1-m perturbations (a 2-m separation in semi-major axis), the point masses drift by each other, 
with the three closest approaches between the centers of any two bodies as 266 m, 365 m, and 400 m.\footnote{We re-examined 
with a time step of $2 \times 10^{-4}$ days any approach of two bodies within 1 km from each other, 
which corresponds to about 15 Hill radii for the largest mass.} 
With a density of 0.5 g/cm$^3$, spherical bodies of these masses would have radii ranging from 49 to 89 m. 
Thus, although these closest approaches would not be collisions, they would still be close enough gravitational encounters to provide significant perturbations, in a range of 4 to 6 Hill radii for the largest mass. 
We consider such a system to be fragile.

\section{Discussion and Implications \label{disc}}
To emphasize how fragile the system is, we can estimate the impulse required to perturb a moonlet's semi-major axis by 1 m, similar to what has been done in \citet{Hedman2020}. 
For nearly circular orbits, the standard orbital perturbation equations can describe the rate of change of semi-major axis over time as \citep{Burns76, Hedman18}
\begin{equation}
\frac{\delta a}{\delta t} = 2 n a \frac{F_p}{F_G}
\end{equation}
where the mean motion $n = \sqrt{GM/a^3} \simeq 1751.7^{\circ}/$day, 
$F_p$ is the azimuthal component of the perturbing force, 
$F_G = GMm/a^2 = n^2am$ is the gravitational force on the moonlet from the planet's center, 
$M$ is the planet's mass, and $m$ is the moonlet's mass. 
The moonlet will thus undergo a semi-major axis change $\delta a$ upon receiving an azimuthal impulse
\begin{equation}
F_p \delta t = \frac{F_G}{2na}\delta a = \frac{1}{2}nm \delta a.
\end{equation}
With the range of masses we use for our low-mass case, 
the impulse required to perturb a moonlet's semi-major axis by 1 m ranges from 
$4.2 \times 10^4$ kg m/s to $2.7 \times 10^5$ kg m/s. 
%A collision with a small piece of interplanetary debris could be sufficient to deliver such an impulse. 
For any collision between a moonlet and interplanetary debris, 
the impact speed would be comparable to the D68 orbit speed $v = na \simeq 24$ km/s. 
Dividing the range of impulses by this orbit speed, we get a range of masses roughly from 2 kg to 10 kg for the interplanetary impactor.
Assuming a density of 0.5 g/cm$^3$, the piece of interplanetary debris would need to be 0.1 m to 0.2 m in radius.  
The estimated cumulative influx rate $\Phi$ for debris of this size is around $10^{-17}/\textrm{m}^2/\textrm{s}$ \citep{Tiscareno13}.
Thus the rough timescale $t \simeq 1/\Phi A$ on which we can expect such a collision, 
using the cross-sectional area $A$ for moonlets with radii of 49 m to 89 m, 
corresponds to a range from 130,000 years to 420,000 years, but this is the impact timescale for just one of the objects. 
Because an impact into any of the objects can break the system, we can adjust the system timescale to about 40,000 years by adding their cross-sectional areas together. 
We therefore cannot expect a co-orbital configuration at D68 to last longer than a few tens of thousands of years. 
For this reason we call the system fragile and find it unlikely that a co-orbital system could explain the orbital evolution of the clumps or the ringlet.

Consequently, we look for other resonances that could drive the orbital evolution of the clumps or the ringlet. 
It is unlikely that a corotation resonance with any satellite is responsible for the clumping of material into ring arcs. 
A 30$^{\circ}$ separation between clumps would be the result of a 12-fold pattern at the D68 semi-major axis. 
A 12-fold pattern could be caused by a 13:12 corotation resonance with an external perturber or an 11:12 corotation resonance with an internal perturber.
A 13:12 corotation resonance with an external perturber would require a perturber at a semi-major axis of 71,300 km, 
which is not as far out as D72, the structure closest to D68.
An 11:12 corotation resonance with an internal perturber would require a perturber at a semi-major axis of 63,800 km,
which is a few thousand km away from Saturn's equatorial radius (60,268 km).
There is no evidence for any moons or ringlets in these regions.   
Moreover, no results came from a numerical search for corotation resonances up to fourth-order between D68 and Janus, Mimas, Enceladus, Tethys, Dione, Rhea, or Titan. 

\begin{table}
\begin{center}
\caption{Predicted corotation resonance locations, 
the $r_{M14}$ column corresponding to our predictions based on the pattern periods reported in \citet{Marley14}, 
the $r_{M19}$ column corresponding to our predictions based on the pattern speeds reported in \citet{Mankovich19}
 \label{tbl-3}}
\begin{tabular}{ccrr}
\tableline
\tableline
$\ell$ & $m$ & $r_{M14}$ (km) & $r_{M19}$ (km) \\
\tableline
8 &6  & 67,852 &67,663 \\
3 &3  & 67,732 &67,932 \\
2 &2  & 66,132 &67,235 \\
\tableline
\end{tabular}
\end{center}
\end{table}

It is possible that a resonance of some sort with Saturn itself could be responsible for the D68 clumping. 
The outer Lindblad resonance of Saturn's $\ell = 5$, $m=3$ oscillation mode is located in the D68 region, 
reported first at 67,625 km $\pm$ 550 km \citep{Marley93} and more recently at roughly 67,550 km \citep{Marley14}.
Although Lindblad resonances do not confine material, 
%they each tend to be accompanied by a nearby corotation resonance, which does confine material. 
each such resonance can be associated with a corotation resonance, which can confine material. 
To locate these corotation resonances, we computed the radii at which the mean motion (using the second-order equation from \citealt{Renner06}) 
matches the pattern speeds associated with the modes reported in \citet{Marley14} and \citet{Mankovich19}.
The modes that produce corotation resonances near D68 are listed in Table \ref{tbl-3}.
Because the pattern speed is dependent on Saturn's structure, 
any of these modes could possibly be responsible for providing a corotation resonance to confine D68 material. 
Mode splitting or mixing could also be involved \citep{Fuller14}, 
allowing the locations of these resonances to fall at slightly different radii than what we can compute.
For a corotation resonance of one of these modes to explain the D68 clumping,
it would require a planet-based angle that moves at a speed comparable to D68's mean motion. 
Although the set of angular separations among the clumps favors a 12-fold pattern,
it is also possible for them to be confined to within the libration longitude of one or a few stable points of a lower $m$ mode, 
and then be spaced out within that external potential. 
Perhaps there is a set of co-orbital moonlets that are trapped together and librating within a larger potential, 
similar to \citet{Renner14}'s model of the Neptune ring arcs. 
Radial oscillations of $\pm$ 10 km have been observed for the D68 ringlet with an estimated period of 14-15 years \citep{Hedman14}, 
though the clumps are drifting more slowly than the rest of the ringlet \citep{Hedman19}.
These radial oscillations could be evidence for that libration. 

In conclusion, 
%although we have been unable to determine a specific factor that dominates the orbital evolution of the D68 ringlet, 
%we can rule out many factors that play roles for other features in the rings. 
%We do find that inasmuch as the dynamics of the D68 clumps can be modeled as a co-orbital system, 
%we predict 
we have tested and ruled out long-term stable co-orbital configurations as an explanation for the spacing of the D68 clumps. 
We therefore predict
that either the clumping is a transient phenomenon,
or that an external mechanism is trapping the clumps in this region. 

\acknowledgements
We are grateful to many individuals for useful discussions, especially P. Ricker for guiding J. A'Hearn's initial studies on co-orbitals; 
B. Sicardy, M. El Moutamid, and M. \'Cuk for insightful questions and clarifications at the 2019 DDA meeting; H. Salo for helpful correspondence, 
and two anonymous reviewers for their helpful comments that have improved this paper. 
We also thank NASA for the support through the Cassini Data Analysis and Participating Scientist Program grant NNX15AQ67G.

\bibliography{Coorbitalsbib} 
\bibliographystyle{aasjournal}

\end{document}